\documentclass[12pt,journal,onecolumn]{IEEEtran}
\usepackage{amsfonts,color,morefloats,pslatex}
\usepackage{amssymb,amsthm, amsmath,latexsym}
\usepackage{makecell}

\usepackage{amsmath,amssymb}
\usepackage{tikz}

\usepackage[hidelinks]{hyperref}
\usepackage{microtype}

\newtheorem{theorem}{Theorem}
\newtheorem{lemma}[theorem]{Lemma}

\newtheorem{example}[theorem]{Example}

\newtheorem{remark}[theorem]{Remark}

\newcommand{\f}{{\mathbb{F}}}

\newcommand{\tr}{{\mathrm{Tr}}}

\newcommand{\wt}{{\mathtt{wt}}}

\newcommand{\F}{{\mathbb{F}}}

\newcommand{\C}{{\mathcal{C}}}

\usepackage{xcolor}

\usepackage{amsmath}
\usepackage{amssymb}

\DeclareMathOperator{\im}{im}

\DeclareMathOperator{\Sub}{Sub}

\usepackage{blindtext}
\usepackage{float}

\ifCLASSINFOpdf

\else

\fi

\UseRawInputEncoding % Arxiv needs this "\UseRawInputEncoding".

\begin{document}

\title{Several families of linear codes via weight functions of three-weight codes   \thanks{
}
}

\author{
Can Xiang  \thanks{C. Xiang is with the College of Mathematics and Informatics, South China Agricultural University, Guangzhou, Guangdong 510642, China (email:cxiangcxiang@hotmail.com). }% <-this % stops a space
and  Chunming Tang \thanks{C. Tang is with the School of Information Science and Technology, Southwest Jiaotong University, Chengdu, Sichuan 610031, China(email: tangchunmingmath@163.com).}
}

\maketitle

\begin{abstract}
Recently, Huang et al. proposed a general framework for deriving new linear codes from a given code by applying weight functions to the set of codewords with a prescribed Hamming weight in [A structural theory of linear codes from weight functions: constructions, geometry and applications, IEEE Trans. Inf. Theory, DOI:10.1109/TIT.2026.3742840, 2026]. While this method was systematically developed for two-weight codes and yielded several families of projective two-weight codes, extending it to codes with more weights remains an interesting and challenging problem. In this paper, we generalize this construction to three-weight codes by employing a well-known family of ternary three-weight cyclic codes.
By determining the weight distribution of every codimension-one subcode, as well as the distribution of fixed-weight codewords across these subcodes, we completely characterize the behavior of the weight-function construction on this family.
As a result, we obtain three  classes of linear codes with four nonzero weights and explicitly determine their parameters. Furthermore, the results show that these obtained codes are self-orthogonal and  minimal.
\end{abstract}


\IEEEpeerreviewmaketitle

\section{Introduction}\label{sec:introduction}
Let $p$ be a prime and $q = p^m$ for some positive integer $m$. Let $\mathbb{F}_q$ be the finite field
of cardinality $q$. An $[n,\, k,\,d]$ linear code $\C$ over $\mathbb{F}_q$ is a $k$-dimensional subspace of $\mathbb{F}_q ^n$ with minimum (Hamming) distance $d$.
Let $A_i$ denote the number of codewords with Hamming weight $i$ in a code
$\C$ of length $n$. The weight enumerator of $\C$ is defined by
$
1+A_1z+A_2z^2+ \cdots + A_n z^n.
$
The sequence $(1,A_1,\ldots,A_n)$ is called the weight distribution of $\C$. A code $\C$ is said to be a $t'$-weight code  if the number of nonzero $A_i$ in the sequence $(A_1, A_2, \cdots, A_n)$ is equal to $t'$. Denote by $\C^\bot$  and $(A_0^{\perp}, A_1^{\perp}, \dots, A_n^{\perp})$ the dual code of a linear code $\C$ and its weight distribution, respectively.
The \emph{Pless power moments} \cite{Huffman2003}, i.e.,
\begin{align}\label{eq:PPM}
 \sum_{i=0}^n  i^t A_i= \sum_{i=0}^t (-1)^i A_i^{\perp}   \left [  \sum_{j=i}^t   j ! S(t,j) q^{k-j} (q-1)^{j-i} \binom{n-i}{n -j}  \right ],
 \end{align}
play an important role in calculating
the weight distributions of linear codes, where $A_0=1$, $0\le t \le n$ and $S(t,j)=\frac{1}{j!} \sum_{i=0}^j  (-1)^{j-i} \binom{j}{i} i^t$.

It is well known that the weight distribution of $\C$ is an important research topic in coding theory, as it contains crucial information about the error correcting capability of the code. Thus the study of the weight distribution has attracted much attention in coding theory and much work focuses on the determination of
the weight distributions of linear codes (see, for example, \cite{Ding-Niederreiter,ding2018,Ding16,sihem2020,sihem2017,zhou20131,Ding2015}).
In particular, many linear codes were produced based on a defining-set
approach and their weight distributions were also determined by using various methods(see, for example, \cite{Ding2015,KDingCDing2015}). In addition, shortening and puncturing techniques are two important approaches to constructing new linear codes. Some linear codes with good parameters have been obtained via these two techniques and their parameters were also determined(see, for example, \cite{liu2021,xiang2022}).

Very recently, Huang et al. \cite{Huang2026} introduced a novel method for constructing linear codes by means of weight functions. More precisely, for a given linear code $\mathcal C$, they consider the set of all codewords of a fixed Hamming weight $w$, and use this set as the coordinate set of a new code by evaluating all linear functionals on $\mathcal C$. Let $L(\mathcal C)$ denote the space of all linear functionals on $\mathcal C$. For each weight $w$ with $A_w \neq 0$, let $\mathrm{wt}^{-1}(w)$ denote the set of all codewords of weight $w$. The new linear code $\mathcal C^*(w)$ is defined as
\begin{equation}\label{eq:C*}
\mathcal C^*(w) = \left\{ \bigl(\ell(\mathbf{c})\bigr)_{\mathbf{c}\in\mathrm{wt}^{-1}(w)} : \ell \in L(\mathcal C) \right\}.
\end{equation}
This construction elevates the set of all $w$-weight codewords from a purely combinatorial object to a defining set for a new code, thereby turning the weight structure of $\mathcal C$ into a mechanism for generating secondary codes.
Within this construction, Huang et al. \cite{Huang2026} established an upper bound on the minimum weight of projective two-weight codes and completely characterized all two-weight codes that achieve this bound, and many new linear codes with excellent parameters were produced. However, their work was exclusively focused on two-weight codes, and the application of this weight-function construction to $t$-weight codes remains unexplored, where $t\geq 3$. A natural question therefore arises: can this construction be extended to $t$-weight codes with $t\geq3$? In fact, some new codes $\mathcal{C}^*(w)$ with good parameters could be produced by properly selecting the code $\mathcal{C}$ under this construction,  but it is in general hard to determine their parameters. To address this problem, we develop a systematic geometric approach that connects the weight distribution of $\mathcal{C}^*(w)$ with the codimension-one subcodes of $\mathcal{C}$. In this paper, we extend this construction method to three-weight codes for constructing new codes and employ a known family of three-weight ternary cyclic codes \cite{ZhouDing2013}
\begin{eqnarray}\label{C}
\C= \left\{ (\tr_{q/p} (  a (-\pi)^t+b \pi^{(p^{k'}+1)t/2} ) )_{t=0}^{q-2} : ~a,b\in \mathbb{F}_{q}        \right\}
\end{eqnarray}
where $p=3$, $q=p^m$, $m\geq3$ is odd and $k'$ is even with $\gcd(m,k')=1$, $\pi$ is a primitive element of the finite field $\mathbb{F}_{q}$ , and $\tr_{q/p}$ is the trace function from $\mathbb{F}_q$  to $\mathbb{F}_p$. By applying this construction to this family of three-weight cyclic codes $\C$, we obtain three new families of linear codes with four nonzero-weights and completely determine their parameters. Moreover, the results of this paper show that these obtained codes are self-orthogonal and minimal.
%Beyond this specific class of cyclic codes, we establish a general analytical framework based on the geometric properties of codimension-one subcodes and dual distance constraints, which is universally applicable to other ternary cyclic codes derived from even planar functions satisfying their dual distance $d^\perp \ge 4$.

%For a codimension-one subcode $\mathcal D$ of $\mathcal C$, let $\mathbf{c}_{\mathcal D}$ denote the codeword in $\mathcal C^*(w)$ corresponding to any linear functional whose kernel is $\mathcal D$.

%\begin{corollary}[Corollary 7, \cite{Huang2026}]Let $\mathcal{C}$ be an $[n,k]$ linear code with $A_w \ne 0$. Let \(\mathcal D\) denote a codimension-one subcode of \(\mathcal C\), i.e., a linear subspace of \(\mathcal C\) with \(\dim \mathcal D = \dim \mathcal C - 1\).If \(\dim \mathcal C^*(w)=\dim \mathcal C\), then the weight distribution of \(\mathcal C^*(w)\) is completely determined by the distribution of \(A_w(\mathcal D)\) over all such \(\mathcal D\).Specifically, for each such \(\mathcal D\), the corresponding codeword $\mathbf{c}_{\mathcal D}$ in \(\mathcal C^*(w)\) has weight\[\wt(\mathbf{c}_{\mathcal D}) = A_w - A_w(\mathcal D).\]\end{corollary}

The rest of this paper is arranged as follows. Section \ref{sec-pre} introduces some notation and results
related to linear codes and their associated projective sets, which will be needed in subsequent sections.
Section \ref{sec-subcodes} gives some properties on codimension-one subcodes of linear codes.
Section \ref{sec-codes} gives some ternary self-orthogonal minimal codes and determines their parameters. Section  \ref{sec-summary} concludes this paper and makes concluding remarks.

\section{Preliminaries}\label{sec-pre}

In this section, we briefly recall some results
related to linear codes and their associated projective sets over finite fields. These results will be used later in this paper.

\subsection{Some results on linear codes}
Let $\C$ be an  $[n,\, k,\,d]$ linear code over $\mathbb{F}_q$. A linear code $\C$ is called a self-orthogonal code if $\C \subseteq  \C^\bot$. There is much work on constructing self-orthogonal codes by various methods (see, for example, \cite{kai2024,Kohnert2009,lirh2024,wwf2026,sihem2025}). Especially, some researchers (see, for example, \cite{heng-dcc,heng-dis}) obtained self-orthogonal codes by  applying the augmented technique to some known linear codes based on the defining set method \cite{Ding2015,KDingCDing2015,Lisihem2020} and judged whether these augmented codes are self-orthogonal by their parameters and Lemma \ref{lem-judge}.

\begin{lemma}\cite{Huffman2003} \label{lem-judge}
Let $\C$ be a linear code over $\mathbb{F}_{p}$. If $p=2$  and every codeword of $\C$  has weight divisible by four, then $\C$  is self-orthogonal. If $p=3$ , $\C$  is self-orthogonal if and only if every codeword of $\C$  has weight divisible by three.
\end{lemma}

An $[n,\, k,\,d]$ linear code $\C$ is said to be {\em cyclic} if
$(c_0,c_1, \cdots, c_{n-1}) \in \C$ implies $(c_{n-1}, c_0, c_1, \cdots, c_{n-2}) \in \C$. Let $\C$ be an $[n,k,d]$ cyclic code over $\mathbb{F}_q$. If $n=q^m-1$ (resp. $n=q^m+1$), the cyclic code $\C$ is called primitive (resp. antiprimitive).
If we identify a vector $(c_0,c_1, \cdots, c_{n-1}) \in \f_q ^n$
with the following polynomial
$$\sum_{i=0}^{n-1} c_ix^i   \in \f_q[x]/(x^n-1),$$
then any cyclic code $\C$ of length $n$ over $\mathbb{F}_q$ is an ideal of the quotient ring $\f_q[x]/(x^n-1)$.
It is known that the ring $\f_q[x]/(x^n-1)$ is a principal ideal ring. Thus, for any
cyclic code $\C$ of length $n$ over $\mathbb{F}_q$, there exists a unique monic divisor $g(x)$ of $x^n-1$ of the smallest degree such that $\C=\langle g(x) \rangle$ . This polynomial $g(x)$ is called the {\em generator polynomial,} and $h(x)=(x^n-1)/g(x)$ is called the {\em check} polynomial of $\C$.
Note that  $k=n-deg(g(x))$  and $\{g(x),x g(x), \cdots ,  x^{k-1} g(x) \}$ is a basis of $\C$. It is obvious that the check polynomial of the cyclic code $\C$ defined in (\ref{C}) is $h_1(x)h_2(x)$, where $h_1(x)$ and $h_2(x)$ are the minimal polynomials of $(-\pi)^{-1}$ and $\pi^{-(p^{k'}+1)/2}$ over $\mathbb{F}_p$, respectively.

\subsection{Linear codes and their associated projective sets}
Let $\mathrm{PG}(k-1, q)$ be the $(k-1)$-dimensional projective space over $\mathbb{F}_q$. Let $\mathcal{C}$ be an $[n, k, d]$ projective linear code over $\mathbb{F}_q$ (i.e., the minimum distance $d(\mathcal{C}^\perp) $ of $\C^\perp $ satisfies $d(\mathcal{C}^\perp) \ge 3$) with a generator matrix $G = [\mathbf{g}_1, \mathbf{g}_2, \dots, \mathbf{g}_n] \in \mathbb{F}_q^{k \times n}$. Consequently, no column is all-zero vector and no two columns are linearly dependent in $G$.
Thus, each non-zero column vector $\mathbf{g}_i \in \mathbb{F}_q^k$ uniquely determines a projective point in $\mathrm{PG}(k-1, q)$ and we define it as the 1-dimensional subspace:
\begin{equation} P_i = [\mathbf{g}_i] = \{\alpha \mathbf{g}_i \mid \alpha \in \mathbb{F}_q^*\}, \quad i = 1, 2, \dots, n.
\end{equation}
This means that the generator matrix $G$ defines an $n$-element projective point set:
\begin{equation}
\mathcal{K} = \{P_1, P_2, \dots, P_n\} \subset \mathrm{PG}(k-1, q).
\end{equation}
The parameters of the code $\mathcal{C}$ and its corresponding point set $\mathcal{K}$ are connected as follows:
\begin{enumerate}
\item The code length is equal to the cardinality of the point set, i.e., $n = |\mathcal{K}|$.
\item For any non-zero row vector $\mathbf{u} \in \mathbb{F}_q^k \setminus \{\mathbf{0}\}$, let $H_{\mathbf{u}}$ denote the projective hyperplane in $\mathrm{PG}(k-1, q)$ defined by: \begin{equation} H_{\mathbf{u}} = \{[\mathbf{x}] \in \mathrm{PG}(k-1, q) \mid \mathbf{u}\mathbf{x} = 0\}, \end{equation} where $\mathbf{x} \in \mathbb{F}_q^k \setminus \{\mathbf{0}\}$ is a column vector. The Hamming weight of the codeword $\mathbf{c} = \mathbf{u}G \in \mathcal{C}$ is determined by the intersection of $H_{\mathbf{u}}$ with $\mathcal{K}$: \begin{equation} \mathrm{wt}(\mathbf{c}) = n - |H_{\mathbf{u}} \cap \mathcal{K}|. \end{equation}
\item The minimum Hamming distance $d$ of $\mathcal{C}$ is given by: \begin{equation} d = n - \max_{H \in \mathcal{H}} |H \cap \mathcal{K}|, \end{equation} where $\mathcal{H}$ denotes the family of all projective hyperplanes in $\mathrm{PG}(k-1, q)$.
\end{enumerate}
Note that two projective linear codes $\mathcal{C}_1$ and $\mathcal{C}_2$ are monomially equivalent if and only if their corresponding projective point sets $\mathcal{K}_1$ and $\mathcal{K}_2$ are projectively equivalent under the action of the projective general linear group $\mathrm{PGL}(k, q)$.

It is well known that $G = [\mathbf{g}_1, \mathbf{g}_2, \dots, \mathbf{g}_n] \in \mathbb{F}_q^{k \times n}$ is called a parity-check matrix of the dual code $\mathcal{C}^\perp$ of $\C$. For any vector $\mathbf{x} = (x_1, x_2, \dots, x_n) \in \mathbb{F}_q^n \setminus \mathcal{C}^\perp$, its non-zero syndrome column vector is given by
\begin{equation}
\mathbf{s}^T = G\mathbf{x}^T = \sum_{i=1}^n x_i \mathbf{g}_i \in \mathbb{F}_q^k \setminus \{\mathbf{0}\}.
\end{equation}
The 1-dimensional subspace spanned by $\mathbf{s}^T$ defines a point in the ambient projective space $\mathrm{PG}(k-1, q)$, which is called the \emph{projective syndrome point} of $\mathbf{x}$, denoted by:
\begin{equation}
[\mathbf{s}] = [G\mathbf{x}^T] = \left[ \sum_{i=1}^n x_i \mathbf{g}_i \right] \in \mathrm{PG}(k-1, q).
\end{equation}
%Geometrically, the point $[\mathbf{s}]$ acts as a normal vector defining a hyperplane $\{\mathbf{u} \in \mathbb{F}_q^k \mid \mathbf{u}\mathbf{s}^T = 0\}$ in the message space, which uniquely determines a codimension-$1$ subcode \begin{equation} \mathcal{C}_{\mathbf{x}} = \{\mathbf{c} \in \mathcal{C} \mid \mathbf{c}\mathbf{x}^T = 0\} = \{\mathbf{u}G \in \mathcal{C} \mid \mathbf{u}\mathbf{s}^T = 0\}. \end{equation} Consequently, there exists a one-to-one correspondence between the codimension-$1$ subcodes of $\mathcal{C}$ and the points in the projective space $\mathrm{PG}(k-1, q)$.

\section{Some properties on codimension-one subcodes of linear codes }\label{sec-subcodes}

In this section, we give some properties on codimension-one subcodes of linear codes.
We first show that each codimension-one subcode of linear code $\C$ corresponds uniquely to a $1$-dimensional extension of its dual code $\mathcal{C}^\perp$, which is described in Lemma \ref{lem:codim1_duality}.

\begin{lemma}\label{lem:codim1_duality}
Let $\mathcal{C}$ be an $[n, k]$ linear code over $\mathbb{F}_q$. If $\mathcal{D} \subseteq \mathcal{C}$ is a codimension-one subcode of $\C$ (i.e., $\dim \mathcal{D} = k - 1$), then there exists a vector $\mathbf{v} \in \mathbb{F}_q^n \setminus \mathcal{C}^\perp$ such that
\begin{equation}
\mathcal{D} = \{ \mathbf{c} \in \mathcal{C} \mid \mathbf{c} \cdot \mathbf{v} = 0 \} = \mathcal{C} \cap \langle \mathbf{v} \rangle^\perp,
\end{equation}
and its dual code $\mathcal{D}^\perp$ is a $1$-dimensional extension of $\mathcal{C}^\perp$, given by
\begin{equation}
\mathcal{D}^\perp = \mathcal{C}^\perp + \langle \mathbf{v} \rangle.
\end{equation}
Conversely, for any $\mathbf{v} \in \mathbb{F}_q^n \setminus \mathcal{C}^\perp$, the subspace $\mathcal{D} = \mathcal{C} \cap \langle \mathbf{v} \rangle^\perp$ is a codimension-one subcode of $\mathcal{C}$. \end{lemma}
\begin{proof}
Since $\mathcal{D}$ is a subspace of $\mathcal{C}$ with $\dim \mathcal{D} = k - 1$, there exists a linear functional $f: \mathcal{C} \to \mathbb{F}_q$ such that $\ker(f) = \mathcal{D}$. By the non-degeneracy of the standard Euclidean inner product on $\mathbb{F}_q^n$, there exists a vector $\mathbf{v} \in \mathbb{F}_q^n$ such that $f(\mathbf{c}) = \mathbf{c} \cdot \mathbf{v}$ for all $\mathbf{c} \in \mathcal{C}$, which implies $\mathcal{D} = \mathcal{C} \cap \langle \mathbf{v} \rangle^\perp$. Because $\mathcal{D} \subsetneq \mathcal{C}$, the functional $f$ is non-zero, meaning $\mathbf{v} \notin \mathcal{C}^\perp$. Applying the duality property of orthogonal subspaces, we have
\begin{equation} \mathcal{D}^\perp = (\mathcal{C} \cap \langle \mathbf{v} \rangle^\perp)^\perp = \mathcal{C}^\perp + (\langle \mathbf{v} \rangle^\perp)^\perp = \mathcal{C}^\perp + \langle \mathbf{v} \rangle.
\end{equation}
Since $\mathbf{v} \notin \mathcal{C}^\perp$, $\dim(\mathcal{D}^\perp) = \dim(\mathcal{C}^\perp) + 1 = (n - k) + 1$. This completes the proof.
\end{proof}

Next we will show that there is a one-to-one correspondence between codimension-$1$ subcodes of $\mathcal{C}$ and the points of the projective space $\mathrm{PG}(k-1, q)$.

\begin{lemma}\label{one-one}
Let $\mathcal{C}$ be an $[n, k]$ linear code over $\mathbb{F}_q$. Let $\mathrm{Sub}_{k-1}(\mathcal{C})$ be the set of all codimension-$1$ subcodes of $\mathcal{C}$. There is a natural one-to-one correspondence between $\mathrm{Sub}_{k-1}(\mathcal{C})$ and the points of the projective space $\mathrm{PG}(k-1, q)$, i.e.,
\[
\mathrm{Sub}_{k-1}(\mathcal{C}) \longleftrightarrow \mathrm{PG}(k-1, q).
\]
Moreover,
\[
|\mathrm{Sub}_{k-1}(\mathcal{C})| = \frac{q^k - 1}{q - 1}.
\]
\end{lemma}

\begin{proof}
Note that every codimension-$1$ subcode $\mathcal{D} \subset \mathcal{C}$ is the kernel of a non-zero linear functional $f \in L(\mathcal C)$, i.e.,
\[
\mathcal{D} = \ker(f) = \{ c \in \mathcal{C} \mid f(c) = 0 \},
\]
where $L(\mathcal C)$ is the dual space of $\C$.
For any two non-zero linear functionals $f, g \in L(\mathcal C)$,  we have $\ker(f) = \ker(g)$ if and only if $g = \lambda f$ for some non-zero scalar $\lambda \in \mathbb{F}_q^*$.
Thus, each codimension-$1$ subspace $\mathcal{D} \in \Sub_{k-1}(\mathcal{C})$ uniquely corresponds to a $1$-dimensional subspace (a projective point)
$[f] = \langle f \rangle$
in the dual space $L(\mathcal C) \cong \mathbb{F}_q^k$.

Explicitly, let $G \in \mathbb{F}_q^{k \times n}$ be a generator matrix of $\mathcal{C}$ whose rows are $\mathbf{r}_1, \dots, \mathbf{r}_k$. Then the isomorphism $L(\mathcal{C}) \cong \mathbb{F}_q^k$ is given by $f \mapsto \mathbf{s} = (f(\mathbf{r}_1), \dots, f(\mathbf{r}_k))^T$, which identifies the line $[f]$ with a unique projective point $P = [\mathbf{s}] \in \mathrm{PG}(k-1, q)$.
This yields the bijection
\[
\Sub_{k-1}(\mathcal{C}) \longleftrightarrow \mathbb{P}(L(\mathcal C)) \cong \mathrm{PG}(k-1, q),~~~~~~\mathcal{D} \longleftrightarrow P = [\mathbf{s}].
\]
Since the total number of non-zero functionals in $L(\mathcal C)$ is $q^k - 1$, and each $1$-dimensional subspace contains $q - 1$ non-zero functionals, the total number of such subcodes is
\[
|\Sub_{k-1}(\mathcal{C})| = \frac{q^k - 1}{q - 1}.
\]
This completes the proof.
\end{proof}

Let $p$ be an odd prime and $q = p^m$. Let $\mathcal{C}$ be an $[n, k]$ linear code over $\mathbb{F}_p$ with a generator matrix $G = (\mathbf{g}_1, \dots, \mathbf{g}_n) \in \mathbb{F}_p^{k \times n}$. Let $S = \{[\mathbf{g}_1], \dots, [\mathbf{g}_n]\} \subset \mathrm{PG}(k-1, p)$ be the projective point set determined by the columns of $G$. Let $\mathcal{D}$ be a codimension-$1$ subcode of $\mathcal{C}$, and denote by $P=[\mathbf{s}] \in \mathrm{PG}(k-1, p)$ the projective syndrome point corresponding to $\mathcal{D}$. Let $r(P)$ be the number of secant lines of $S$ passing through $P$ (with the standard convention that $r(P) = 0$ if no such secant line exists). Then we have the following result in Lemma \ref{3case}, which plays a crucial role in determining the parameters of $\mathcal{D}$.

\begin{lemma}\label{3case}
If $d(\mathcal{C}^\perp) \geq 4$, then the value of  $(A_1(\mathcal{D}^\perp), A_2(\mathcal{D}^\perp))$  has the following two cases:
\begin{description}
\item[\textbf{Case I:}] ~~If $P \in S$, then $A_1(\mathcal{D}^\perp) = p-1$ and $A_2(\mathcal{D}^\perp) = 0$.
\item[\textbf{Case II:}]~~~If $P \notin S$, then $A_1(\mathcal{D}^\perp) = 0$ and $A_2(\mathcal{D}^\perp) = (p-1)r(P)$.  In particular, $A_2(\mathcal{D}^\perp) = 0$ if and only if $P$ does not lie on any secant line of $S$.
\end{description}
\end{lemma}

\begin{proof}
Note that the corresponding projective syndrome point of $\mathcal{D}$ is $P=[\mathbf{s}]$ with $\mathbf{s} \in \mathbb{F}_p^k \setminus \{\mathbf{0}\}$. The quotient space $\mathcal{D}^\perp / \mathcal{C}^\perp$ has $p-1$ non-zero cosets, which are in one-to-one correspondence with the non-zero syndromes $\{\lambda \mathbf{s} : \lambda \in \mathbb{F}_p^*\}$ (i.e., the cosets $\mathbf{v}_\lambda + \mathcal{C}^\perp$, where $G\mathbf{v}_\lambda^T = \lambda \mathbf{s}$).
\begin{description}
\item[\textbf{Case I:}] ~Since $d(\mathcal{C}^\perp) \geq 4$, each column $\mathbf{g_i}$ in $G$ satisfies $\mathbf{g_i} \neq \mathbf{0}$. For any codeword with $1$-weight in $\mathcal{D}^\perp$, its syndrome must be a non-zero scalar multiple of some column of the generator matrix $G$. Thus, $A_1(\mathcal{D}^\perp) >0$ if and only if $P \in S$. Suppose that there are two distinct codewords with $1$-weight in the same coset, then $d(\mathcal{C}^\perp) \leq 2$, which contradicts the assumption that $d(\mathcal{C}^\perp) \geq 4$. Therefore, each non-zero coset contains at most one codeword with $1$-weight. When $P \in S$, there is a unique column $\mathbf{g}_i$ such that $[\mathbf{g}_i] = P$, meaning that for each $\lambda \in \mathbb{F}_p^*$, the coset corresponding to syndrome $\lambda \mathbf{s}$ contains exactly one codeword with $1$-weight. Hence, there are precisely $|\mathbb{F}_p^*| = p-1$ such codewords, which gives $A_1(\mathcal{D}^\perp) = p-1$.

Furthermore, when $A_1(\mathcal{D}^\perp)> 0$, if $A_2(\mathcal{D}^\perp)\neq 0$, there would exist a codeword of $2$-weight in $\mathcal{D}^\perp$, belonging to some coset with syndrome $\lambda \mathbf{s}$. Combining the $1$-weight codeword in the same coset, this will give a non-zero codeword with weight at most $1+2=3$ in $\mathcal{C}^\perp$. Thus, $d(\mathcal{C}^\perp) \leq 3$, which contradicts the assumption that $d(\mathcal{C}^\perp) \geq 4$. Thus, $A_2(\mathcal{D}^\perp) = 0$.

\item[\textbf{Case II:}]~~When $P \notin S$, from the proof of  the above Case I we have $A_1(\mathcal{D}^\perp) = 0$. If there are no $2$-weight codewords in $\mathcal{D}^\perp$ (i.e., $A_2(\mathcal{D}^\perp) = 0$), then $P$ cannot lie on any secant line spanned by pairs of points in $S$, which matches $r(P) = 0$.
Otherwise, each $2$-weight codeword $\mathbf{c} \in \mathcal{D}^\perp$ satisfies
$$G \mathbf{c}^T =a\mathbf{g}_i + b\mathbf{g}_j =\lambda \mathbf{s}$$

with $a,b,\lambda  \in \mathbb{F}_p^*$ for some distinct indices $i, j \in \{1, 2, \dots, n\}$, which means that the secant line $\langle [\mathbf{g}_i], [\mathbf{g}_j] \rangle$ passes through the projective syndrome point $P=[\mathbf{s}]$. Conversely, since $d(\mathcal{C}^\perp) \geq 4$, any three columns of $G$ are linearly independent. Thus, each secant line through $P$ uniquely determines the index pair $\{i, j\}$. For each fixed $\lambda \in \mathbb{F}_p^*$, the linear system $a\mathbf{g}_i + b\mathbf{g}_j = \lambda \mathbf{s}$ has a unique solution for $(a, b) \in (\mathbb{F}_p^*)^2$. Thus, each secant line contributes exactly $|\mathbb{F}_p^*| = p-1$ codewords of $2$-weight to $\mathcal{D}^\perp$. Consequently, $A_2(\mathcal{D}^\perp) = (p-1)r(P)$ holds for all $r(P) \geq 0$.
\end{description}
\end{proof}

\section{Linear codes from weight functions of three-weight cyclic codes  }\label{sec-codes}

In this section, the objective is to determine the parameters of the ternary code $\mathcal C^*(w)$ from the cyclic code $\C$ and show that these codes are self-orthogonal and minimal, where $\mathcal C^*(w)$  and  $\C$ were defined in (\ref{eq:C*}) and (\ref{C}).

For convenience, we assume that $p=3$, \(q=3^m\), $\pi$ is a primitive element of $\f_q$, $m\geq 3$ is odd and $k'\geq 2$ is even with \(\gcd(m,k')=1\) throughout this section unless otherwise stated.

Recall that the ternary cyclic code $\C$ defined by (\ref{C}) is as follows:
\begin{equation}\label{cyclic}
\mathcal C = \{ \mathbf c(a,b) : a,b\in\mathbb F_q \},
\end{equation}
where
\begin{equation}\label{cyclic-ab}
\mathbf c(a,b) = \left( \mathrm{Tr}_{q/3}\left(a(-\pi)^t + b\pi^{ut}\right) \right)_{t=0}^{q-2}, ~~u=\frac{3^{k'}+1}{2}.
\end{equation}
%with $u=\frac{3^k+1}{2}.$
The parameters of this cyclic code \(\mathcal C\) were given in Theorem 3.7 of \cite{ZhouDing2013}, i.e., the code \(\mathcal C\)  is an $[n, k]$ cyclic code with three nonzero weights,
where
\begin{equation}\label{eq1}
n=q-1=3^m-1,\quad k=2m,
\end{equation}
and the three nonzero weights are
\begin{equation}\label{eq2}
%w_1=2U-T,\qquad w_2=2U,\qquad w_3=2U+T,
w_1 = 2 \cdot 3^{m-1} - 3^{(m-1)/2}, \qquad w_2 = 2 \cdot 3^{m-1}, \qquad w_3 = 2 \cdot 3^{m-1} + 3^{(m-1)/2}
\end{equation}
and the numbers of their corresponding codewords with $w_i$-weight are
\begin{equation}\label{eq3}
A_{w_1}= \left(3^{m-1} + 3^{(m-1)/2}\right)\left(3^m - 1\right),  A_{w_2} = \left(3^{m-1} + 1\right)\left(3^m - 1\right),   A_{w_3} = \left(3^{m-1} - 3^{(m-1)/2}\right)\left(3^m - 1\right).
\end{equation}
%with\begin{equation}\label{eq4} U=3^{m-1},\qquad T=3^{(m-1)/2}.\end{equation}
Note that the dual code \(\mathcal C^\perp\) is a \([3^m-1, 3^m-1-2m, 4]\) optimal ternary code \cite{ZhouDing2013}. Hence this ternary cyclic code \(\mathcal C\) is projective.

\subsection{Codimension-one subcodes of cyclic codes and their weight distributions}

In this subsection, we will determine the weight distributions of the codimension-one subcodes of the ternary cyclic code $\C$. To this end, we need the following results in Lemma \ref{lem:planarity} and Theorem \ref{thm:secant-distribution}.

Recall that $G = [\mathbf{g}_1, \mathbf{g}_2, \dots, \mathbf{g}_n]$ is a generator matrix of $\C$. By the definition of $\C$ in (\ref{cyclic}), the $t$-th column $\mathbf{g}_t$ of $G$ can be represented as $$((-\pi)^t, \pi^{ut})^\top$$
in $\mathbb{F}_q^2$ for each $t\in \{1,2,...,n\}$.
Multiplying each column $\mathbf{g}_t$ by the non-zero scalar $(-1)^t \in \mathbb{F}_3^*$, we get
$$(-1)^t \cdot ((-\pi)^t, \pi^{ut})^\top=(\pi^t, \pi^{(\frac{q-1}{2}+u)t})^\top .$$
Setting $x = \pi^t \in \mathbb{F}_q^*$, each normalized column vector can be written as the coordinate form
\begin{equation} h(x) = (x, x^e), \quad x \in \mathbb{F}_q^*,
\end{equation}
where
\begin{equation}
 e \equiv\frac{q-1}{2} + u= \frac{q-1}{2} +\frac{3^{k'}+1}{2} =\frac{3^m+3^{k'}}{2} \pmod{q-1}.
 \end{equation}
%Let $f(x) = x^e$. Regard $\mathbb{F}_q^2$ as the $2m$-dimensional vector space $\mathbb{F}_3^{2m}$ over $\mathbb{F}_3$. Then the resulting projective point set in $\mathrm{PG}(2m-1, 3)$ can be expressed as\begin{equation} \label{SS} S = \{ [h(x)] : x \in \mathbb{F}_q^* \} = \{ [(x, f(x))] : x \in \mathbb{F}_q^* \}.\end{equation}
Let $f(x) = x^e$. By fixing a basis of $\mathbb{F}_q$ over $\mathbb{F}_3$, we naturally regard $\mathbb{F}_q^2$ as the $2m$-dimensional vector space $\mathbb{F}_3^{2m}$ over $\mathbb{F}_3$.
Under this identification,
%the ternary coordinates of the code $\C$ in (\ref{cyclic}) are exactly preserved by the trace representation,
the resulting projective point set in $\mathrm{PG}(2m-1, 3)$ can be expressed as
\begin{equation} \label{SS}
    S = \{ [h(x)] : x \in \mathbb{F}_q^* \} = \{ [(x, f(x))] : x \in \mathbb{F}_q^* \}.
\end{equation}
%It is clear that the cyclic code $\mathcal C$ is monomially equivalent to the trace code defined by $f(x)$:\begin{equation} \label{C_f}\mathcal C_f = \left\{ \mathbf c_f(a,b) = \left( \mathrm{Tr}_{q/3}\big( ax + b f(x) \big) \right)_{x \in \mathbb F_q^*} : a, b \in \mathbb F_q \right\}.\end{equation}

To investigate the geometric structure of $S$, we first establish the differential and algebraic properties of the monomial function $f(x)$ in Lemma \ref{lem:planarity}.

\begin{lemma}\label{lem:planarity}
Let the symbols and notations be as above.
Then the monomial function $f(x) = x^e$ is a \emph{planar function} over $\mathbb{F}_q$, i.e., the difference mapping
\begin{equation}
    D_A f(x) = f(x+A) - f(x)
\end{equation}
is a permutation of $\mathbb{F}_q$ for every $A \in \mathbb{F}_q^*$. Furthermore, $f(x)$ satisfies the following properties:
\begin{enumerate}
    \item[\textup{(i)}] $f(-x) = f(x)$ for any $x \in \mathbb{F}_q$ ;
    \item[\textup{(ii)}] $\gcd(e, q-1) = 2$, which implies that $f$ is a $2$-to-$1$ mapping from $\mathbb{F}_q^*$ onto $SQ$, where $SQ $ is the set of all non-zero quadratic residues (squares) in $\mathbb{F}_q$.
    \item[\textup{(iii)}] For any $a \in \mathbb{F}_q^*$ and $b \in \mathbb{F}_q$, the system of equations
    \begin{equation*}
    \begin{cases} x + y = a, \\ f(x) + f(y) = b
    \end{cases}
    \end{equation*}
    has at most one unordered solution $\{x, y\}$ in $\mathbb{F}_q$.
\end{enumerate}
\end{lemma}

\begin{proof} First, we prove that $f(x) = x^e$ is a planar polynomial over $\mathbb{F}_q$. Let $s$ be the unique integer such that $s \equiv m - k' \pmod{2m}$ with $1 \le s < 2m$. Since $k' + s \equiv m \pmod{2m}$, we have $3^{k'+s} \equiv 3^m \pmod{3^{2m}-1}$. Noticing that $2(3^m-1) \mid (3^{2m}-1)$, this implies $3^{k'+s} \equiv 3^m \pmod{2(3^m-1)}$, which yields \begin{equation*} e = \frac{3^m + 3^{k'}}{2} \equiv \frac{3^{k'+s} + 3^{k'}}{2} = 3^{k'} \left( \frac{3^s+1}{2} \right) \pmod{3^m-1}.
\end{equation*}
Since $m$ is odd and $k'$ is even with $\gcd(m, k') = 1$, $s$ is odd and $\gcd(s, 2m) = 1$. By Theorem 4.1 of \cite{Coulter1997}, $g(x) = x^{(3^s+1)/2}$ is a planar polynomial over $\mathbb{F}_q$. Since $x \mapsto x^{3^{k'}}$ is a field automorphism of $\mathbb{F}_q$, as functions on $\mathbb{F}_q$, the function $f(x) = x^e = (g(x))^{3^{k'}}$ is also planar over $\mathbb{F}_q$.

(i) Since $m$ is odd and $k'$ is even, $3^m + 3^{k'} \equiv (-1)^m + (-1)^{k'} \equiv 0 \pmod 4$, which implies that the exponent $e = (3^m + 3^{k'})/2$ is even. Thus, $f(-x) = (-x)^e = x^e = f(x)$ holds for all $x \in \mathbb{F}_q$.

(ii) By Proposition 2.4 of \cite{Coulter1997}, any planar monomial $x^u$ over $\mathbb{F}_q$ satisfies $\gcd(u, q-1) = 2$. Applying this gives $\gcd(e, q-1) = 2$, which implies that the multiplicative homomorphism $x \mapsto x^e$ on $\mathbb{F}_q^*$ has a kernel of size $2$, making it a $2$-to-$1$ map with image size $(q-1)/2$.
Since $e$ is even, every non-zero image is a square, which implies that the image of $\mathbb{F}_q^*$ is precisely $SQ$.

(iii) Substituting $y = a - x$ into $f(x) + f(y) = b$ yields \begin{equation}\label{eq:sum_eq} f(x) + f(a - x) = b. \end{equation} Notice that if $x_0$ is a solution to \eqref{eq:sum_eq}, then $a - x_0$ is also a solution, and both correspond to the same unordered pair $\{x_0, a - x_0\}$. Thus, it suffices to prove that \eqref{eq:sum_eq} has at most two solutions in $\mathbb{F}_q$. Suppose $x_1, x_2 \in \mathbb{F}_q$ are two solutions to \eqref{eq:sum_eq}. Then \begin{equation*} f(x_1) - f(x_2) = f(a - x_2) - f(a - x_1). \end{equation*} Setting $y_1 = a - x_1$, $y_2 = a - x_2$, and $\Delta = x_1 - x_2 = y_2 - y_1$, the above equation becomes \begin{equation*} f(x_2 + \Delta) - f(x_2) = f(y_1 + \Delta) - f(y_1), \end{equation*} that is, $D_\Delta f(x_2) = D_\Delta f(y_1)$. If $\Delta \neq 0$, the planarity of $f$ implies that $D_\Delta f$ is a permutation, forcing $x_2 = y_1 = a - x_1$. If $\Delta = 0$, then $x_2 = x_1$. Consequently, \eqref{eq:sum_eq} has at most two solutions, which corresponds to at most one unordered solution $\{x, y\}$. \end{proof}

By Lemma \ref{lem:planarity}, $f(-x) = f(x) \neq -f(x)$ for any $x \in \mathbb{F}_q^*$. Thus, no two distinct elements in $\{h(x) : x \in \mathbb{F}_q^*\}$ are $\mathbb{F}_3$-linearly dependent. Hence, the mapping $x \mapsto [h(x)]$ is injective, and the projective point set $\mathcal{S}$ has cardinality $|\mathcal{S}| = q-1$. Furthermore, since the dual code $\mathcal{C}^\perp$ has minimum distance $d^\perp = 4$, no three points in $\mathcal{S}$ are collinear, which implies that $\mathcal{S}$ is a $(q-1)$-cap in $\mathrm{PG}(2m-1, 3)$.
Consequently, every secant line intersects $\mathcal{S}$ in exactly two points. For any external point $P \in \mathrm{PG}(2m-1, 3) \setminus \mathcal{S}$, let $r(P)$ denote the number of secant lines of $\mathcal{S}$ passing through $P$. Next we will analyze the distribution of $r(P)$ in Theorem \ref{thm:secant-distribution}, which plays a crucial role in determining the weight distributions of the codimension-one subcodes of the cyclic code $\mathcal{C}$.

\begin{theorem}\label{thm:secant-distribution}
Let the symbols and notations be as above. Let $\mathcal{S} \subset \mathrm{PG}(2m-1, 3)$ be the projective point set defined in (\ref{SS}).
For any external point $P = [(A, B)]_{\mathbb{F}_3} \in \mathrm{PG}(2m-1, 3) \setminus \mathcal{S}$, we have
\begin{equation}
r(P) \in \{1, 2, 3\}.
\end{equation}
%In particular, $r(P) \neq 0$ for all external points, implying that the secant lines of $\mathcal{S}$ completely cover the complement $\mathrm{PG}(2m-1, 3) \setminus \mathcal{S}$.
More precisely:
\begin{enumerate}
\item If $A = 0$, then $B \neq 0$ and $r(P) = 1$;
\item If $A \neq 0$, then $r(P) \in \{1, 2, 3\}$.
\end{enumerate}
\end{theorem}
\begin{proof}
By definition, the projective point $P = [(A, B)]_{\mathbb{F}_3}$ lies on the secant line passing through two distinct points $[h(x)]_{\mathbb{F}_3}, [h(y)]_{\mathbb{F}_3} \in \mathcal{S}$ if and only if $(A, B)$ is an $\mathbb{F}_3$-linear combination of $h(x)=(x, x^e)$ and $h(y)=(y, y^e)$, i.e., this is equivalent to
\begin{equation*} h(x) \pm h(y)=((x, x^e) \pm (y, y^e)) \in \{ (A, B), -(A, B) \}
\end{equation*}
since $\mathbb{F}_3^* = \{1, -1\}$.   Thus, $r(P)$ equals the number of the sets $\{x, y\} \subset \mathbb{F}_q^*$ satisfying either the system (\ref{AA}) or the system (\ref{BB}):
\begin{align} \label{AA}
 & \quad x - y = \varepsilon  A, \quad x^e - y^e = \varepsilon B
\end{align}
\begin{align}\label{BB}
& \quad x + y = \varepsilon A, \quad x^e + y^e = \varepsilon B
\end{align}
where $\varepsilon \in \{1,-1\}$.
We first claim that systems (\ref{AA}) and (\ref{BB}) share no common solution in $\mathbb{F}_q^*$. If a pair $\{x, y\}$ satisfies $x - y = \varepsilon_1 A$ and $x + y = \varepsilon_2 A$ for some $\varepsilon_1, \varepsilon_2 \in \{1, -1\}$, then these equations give $2x = (\varepsilon_2 + \varepsilon_1)A$ and $2y = (\varepsilon_2 - \varepsilon_1)A$. Since $\varepsilon_1, \varepsilon_2 \in \{1, -1\}$, at least one of $\varepsilon_2 \pm \varepsilon_1$ is zero, which forces $x = 0$ or $y = 0$, contradicting the requirement that $\{x, y\} \subset \mathbb{F}_q^*$.

We distinguish two cases as follows:
\begin{itemize}
\item The case that $A = 0$.

In this case, it is clear that there is no 2-subset $\{x, y\} \subset \mathbb{F}_q^*$ satisfying the system (\ref{AA}). From (i) and (ii) in Lemma \ref{lem:planarity}, we have that $e$ is even and $\gcd(e,q-1)=2$. Thus, from (\ref{BB}) we get
\begin{equation*}
x^e + (-x)^e = 2x^e = -x^e = \varepsilon B \implies x^e = -\varepsilon B.
\end{equation*}
Note that $-1$ is a non-square in $\mathbb{F}_q$ since $q = 3^m$ and $m$ is odd. Thus, for any $B \in \mathbb{F}_q^*$, exactly one choice of $\varepsilon \in \{1, -1\}$ makes $-\varepsilon B$ a non-zero square in $\mathbb{F}_q$. For this unique $\varepsilon$, the equation $x^e = -\varepsilon B$ has exactly two opposite solutions in $\mathbb{F}_q^*$, say $\pm x_0$, which yield a unique 2-subset $\{x_0, -x_0\}$. For the other choice of $\varepsilon$, there is no solution. Consequently, there is exactly one 2-subset $\{x, y\} \subset \mathbb{F}_q^*$ satisfying the system (\ref{BB}). Hence, we conclude that $r(P) = 0 + 1 = 1$.

\item The case that $A \neq 0$.

For system (\ref{AA}), substituting $x = y + \varepsilon A$ into the second equation gives \begin{equation*} (y + \varepsilon A)^e - y^e = \varepsilon B \iff D_{\varepsilon A}f(y) = \varepsilon B. \end{equation*} Since $A \neq 0$ and $\varepsilon \in \{1, -1\}$, we have $\varepsilon A \in \mathbb{F}_q^*$. By Lemma~\ref{lem:planarity}, the difference mapping $D_{\varepsilon A}f$ is a permutation of $\mathbb{F}_q$. Thus, for each fixed $\varepsilon$, there is a unique ordered pair $(x, y) \in \mathbb{F}_q \times \mathbb{F}_q$ satisfying (\ref{AA}). If $(x_0, y_0)$ denotes the unique solution for $\varepsilon = 1$, then $(y_0, x_0)$ is the unique solution for $\varepsilon = -1$, both giving the same unordered pair $\{x_0, y_0\}$. Furthermore, if $x_0 = 0$ or $y_0 = 0$, we have $[(A, B)]_{\mathbb{F}_3} = [(\pm y_0, \pm y_0^e)]_{\mathbb{F}_3} \in \mathcal{S}$, contradicting $P \notin \mathcal{S}$. Hence, $x_0, y_0 \in \mathbb{F}_q^*$ with $x_0 \neq y_0$. Therefore, system (\ref{AA}) contributes precisely one $2$-subset $\{x_0, y_0\} \subset \mathbb{F}_q^*$. For system (\ref{BB}), by Lemma~\ref{lem:planarity}~(iii), the system $x + y = \varepsilon A, x^e + y^e = \varepsilon B$ has at most one unordered solution $\{x, y\}$ in $\mathbb{F}_q$ for each $\varepsilon \in \{1, -1\}$. Note that any such solution cannot have $x = y$ or contain $0$, as that would similarly imply $P \in \mathcal{S}$. Furthermore, since $A \neq -A$, the solutions for $\varepsilon = 1$ and $\varepsilon = -1$ are distinct. Therefore, the number $N_{\mathrm{sum}}(P)$ of $2$-subsets in $\mathbb{F}_q^*$ contributed by (\ref{BB}) satisfies $N_{\mathrm{sum}}(P) \in \{0, 1, 2\}$.
Since (\ref{AA}) and (\ref{BB}) share no common solution, summing their contributions yields $r(P) = 1 + N_{\mathrm{sum}}(P) \in \{1, 2, 3\}$.

\end{itemize}
This completes the proof.
\end{proof}

\begin{remark}
The proof of Theorem \ref{thm:secant-distribution} relies only on the planarity of $f(x)=x^e$, the parity of $e$, and the condition $\gcd(e, 3^m-1) = 2$. Consequently,
the results in Theorem~\ref{thm:secant-distribution} can be generalized to any planar monomial $f(x) = x^e$ over $\mathbb{F}_{3^m}$, where $m \ge 3$ is odd and $e$ is an even integer satisfying $\gcd(e, 3^m-1) = 2$.
\end{remark}

%\begin{remark} It is worth mentioning that Theorem \ref{thm:secant-distribution} holds for any even planar function $f(x)$ satisfying $f(0) = 0$ whose non-zero values are all squares in $\mathbb{F}_q$. \end{remark}

Next we will determine the parameters of the codimension-one subcodes of the ternary cyclic code $\C$, which are described in Theorem \ref{thm:sub-distribution}.

\begin{theorem}\label{thm:sub-distribution}
Let the symbols and notations be as above. Let \(\mathcal D\) be a codimension-one subcode of the cyclic code \(\mathcal C\) defined by (\ref{cyclic}). Then the code \(\mathcal D\) is a $[q-1,2m-1]$ linear code with weight distribution having four cases in Tables \ref{tab-weight1}, \ref{tab-weight21}, \ref{tab-weight22} and \ref{tab-weight23}.

\begin{table}[H]
\begin{center}
\caption{Type 1: The weight distribution of $\mathcal D$ when $(A_1(\mathcal{D}^\perp), A_2(\mathcal{D}^\perp))=(2,0)$}\label{tab-weight1}
\begin{tabular}{cc} \hline
Weight  &  Multiplicity   \\ \hline
$0$          &  $1$ \\  [2mm]
$2 \cdot 3^{m-1} - 3^{(m-1)/2}$    &  $ 3^{(m-1)/2}(3^{(3m-3)/2} +2\cdot 3^{m-1} -1           ) $  \\ [2mm]
$2 \cdot 3^{m-1}$                    &  $3^{2m-2}  - 1$                                                        \\ [2mm]
$2 \cdot 3^{m-1} + 3^{(m-1)/2}$    & $ 3^{(m-1)/2}(3^{(3m-3)/2} -2\cdot 3^{m-1} +1           ) $ \\ [2mm]
\hline
\end{tabular}
\end{center}
\end{table}

\begin{table}[H]
\begin{center}
\caption{Type 2: The weight distribution of $\mathcal D$ when $(A_1(\mathcal{D}^\perp), A_2(\mathcal{D}^\perp))=(0,2)$ }\label{tab-weight21}
\begin{tabular}{cc} \hline
Weight  &  Multiplicity   \\ \hline
$0$          &  $1$ \\  [2mm]
$2 \cdot 3^{m-1} - 3^{(m-1)/2}$    & $3^{(m-1)/2} \left( 3^{m-1} - 1 \right) (3^{\frac{m-1}{2}}+1)$   \\ [2mm]
$2 \cdot 3^{m-1}$                    &  $3^{m-1} \left( 3^{m-1} + 2 \right) - 1$                                                        \\ [2mm]
$2 \cdot 3^{m-1} + 3^{(m-1)/2}$    &     $3^{(m-1)/2} \left( 3^{m-1} - 1 \right) (3^{\frac{m-1}{2}}-1)$                                               \\ [2mm]
\hline
\end{tabular}
\end{center}
\end{table}

\begin{table}[H]
\begin{center}
\caption{Type 3: The weight distribution of $\mathcal D$ when $(A_1(\mathcal{D}^\perp), A_2(\mathcal{D}^\perp))=(0,4)$ }\label{tab-weight22}
\begin{tabular}{cc} \hline
Weight  &  Multiplicity   \\ \hline
$0$          &  $1$ \\  [2mm]
$2 \cdot 3^{m-1} - 3^{(m-1)/2}$       & $3^{m-2} \left( 3^m - 1 \right) + 3^{\frac{m-1}{2}}(3^{m-1}-1)$      \\ [2mm]
$2 \cdot 3^{m-1}$                    &  $3^{m-2} \left( 3^m + 2 \right) - 1$                                  \\ [2mm]
$2 \cdot 3^{m-1} + 3^{(m-1)/2}$    & $3^{m-2} \left( 3^m - 1 \right) + 3^{\frac{m-1}{2}}(-3^{m-1}+1)$    \\ [2mm]
\hline
\end{tabular}
\end{center}
\end{table}

\begin{table}[H]
\begin{center}
\caption{Type 4: The weight distribution of $\mathcal D$ when $(A_1(\mathcal{D}^\perp), A_2(\mathcal{D}^\perp))=(0,6)$ }\label{tab-weight23}
\begin{tabular}{cc} \hline
Weight  &  Multiplicity   \\ \hline
$0$          &  $1$ \\  [2mm]
$2 \cdot 3^{m-1} - 3^{(m-1)/2}$    & $3^{m-2} \left( 3^{m} + 1 \right) + 3^{\frac{m-1}{2}}(3^{m-1} -1)$  \\ [2mm]
$2 \cdot 3^{m-1}$                    &  $3^{m-2} \left( 3^{m} - 2 \right) - 1$                                \\ [2mm]
$2 \cdot 3^{m-1} + 3^{(m-1)/2}$    &   $3^{m-2} \left( 3^{m} + 1 \right) + 3^{\frac{m-1}{2}}(1- 3^{m-1} )$                             \\ [2mm]
\hline
\end{tabular}
\end{center}
\end{table}

\end{theorem}

\begin{proof}
Since $\C$ is a $[q-1,2m]$ ternary cyclic code with three nonzero weights in (\ref{eq2}), from the definitions we get that the codimension-one subcode \(\mathcal D\) is a  $[q-1,2m-1]$ ternary linear code with at most three nonzero weights as follows:
\begin{equation}\label{eq:weights}
w_1 = 2 \cdot 3^{m-1} - 3^{(m-1)/2}, \qquad w_2 = 2 \cdot 3^{m-1}, \qquad w_3 = 2 \cdot 3^{m-1} + 3^{(m-1)/2}.
\end{equation}
Note that $d(\mathcal{C}^\perp) = 4$. By Lemma \ref{3case} and Theorem \ref{thm:secant-distribution}, we have
 \begin{equation}\label{eq:dual3}
 (A_1(\mathcal{D}^\perp), A_2(\mathcal{D}^\perp)) \in \{(2,0),~(0,2),~(0,4),~(0,6)\}
 \end{equation}
From (\ref{eq:weights}) and (\ref{eq:dual3}), the first three Pless power moments of (\ref{eq:PPM}) yield the weight distribution in Tables \ref{tab-weight1}, \ref{tab-weight21}, \ref{tab-weight22} and \ref{tab-weight23}.
\end{proof}

Next we will give the  distribution of \(A_{w_i}(\mathcal D)\) over all codimension-one subcodes of \(\mathcal C\), where $w_i$ are described in (\ref{eq:weights}).

\begin{theorem}\label{thm:subcode-distribution-corrected}
Let $m \ge 3$ be an odd integer, $q=3^m$, $n = 3^m - 1$, and let $\mathcal{C}$ be the ternary $[n, 2m]$ cyclic code defined by (\ref{cyclic}). When $\mathcal{D}$ runs through all $\frac{3^{2m}-1}{2}$ codimension-one subcodes of $\mathcal{C}$, the distribution of $A_{w_i}(\mathcal{D})$ ($i \in \{1, 2, 3\}$) is partitioned into four distinct types, whose values and frequencies are given in Table~\ref{tab:subcode-freq-all}.

\begin{table}[H]
\centering \caption{The frequency distribution of $A_{w_i}(\mathcal{D})$ over all codimension-one subcodes of $\mathcal{C}$}\label{tab:subcode-freq-all} \renewcommand{\arraystretch}{1.6}
\begin{tabular}{c c c c c} \hline
Type & $A_{w_1}(\mathcal{D})$ & $A_{w_2}(\mathcal{D})$ & $A_{w_3}(\mathcal{D})$ & Number of $\mathcal{D}$ \\ \hline
1 & $3^{2m-2} + 3^{\frac{m-1}{2}} \left( 2 \cdot 3^{m-1} - 1 \right)$ & $3^{2m-2} - 1$ & $3^{2m-2} - 3^{\frac{m-1}{2}} \left( 2 \cdot 3^{m-1} - 1 \right)$ & $3^m - 1$ \\
2 & $3^{m-1} \left( 3^{m-1} - 1 \right) + 3^{\frac{m-1}{2}} \left( 3^{m-1} - 1 \right)$ & $3^{m-1} \left( 3^{m-1} + 2 \right) - 1$ & $3^{m-1} \left( 3^{m-1} - 1 \right) - 3^{\frac{m-1}{2}} \left( 3^{m-1} - 1 \right)$ & $\frac{(3^m-1)(3^m + 5)}{8}$ \\
3 & $3^{m-2} \left( 3^m - 1 \right) + 3^{\frac{m-1}{2}} \left( 3^{m-1} - 1 \right)$ & $3^{m-2} \left( 3^m + 2 \right) - 1$ & $3^{m-2} \left( 3^m - 1 \right) - 3^{\frac{m-1}{2}} \left( 3^{m-1} - 1 \right)$ & $\frac{3(3^m-1)(3^{m-1} - 1)}{4}$ \\
4 & $3^{m-2} \left( 3^m + 1 \right) + 3^{\frac{m-1}{2}} \left( 3^{m-1} - 1 \right)$ & $3^{m-2} \left( 3^m - 2 \right) - 1$ & $3^{m-2} \left( 3^m + 1 \right) - 3^{\frac{m-1}{2}} \left( 3^{m-1} - 1 \right)$ & $\frac{(3^m-1)(3^m - 3)}{8}$ \\ \hline
\end{tabular}
\end{table}
\end{theorem}

\begin{proof}
Every codimension-one subcode $\mathcal{D}$ of $\mathcal{C}$ corresponds uniquely to a projective point $P \in \mathrm{PG}(2m-1, 3)$.
Since $d(\C^\perp)=4$, the point set $\mathcal{S}$ of $\mathcal{C}$ can be viewed as a cap in $\mathrm{PG}(2m-1, 3)$, which contains $n = q - 1 = 3^m - 1$ points.
%Let $\mathcal{S}$ denote the point set of $\mathcal{C}$. Then $|\mathcal{S}| = q-1$.
By Lemma~\ref{3case} and Theorems~\ref{thm:secant-distribution} and \ref{thm:sub-distribution}, all codimension-one subcodes of $\C$ are partitioned into four types:
%Type 1 corresponds to $P \in \mathcal{S}$ with $N_1 = |\mathcal{S}| = q-1$, while Types 2, 3, and 4 correspond to $P \in \mathrm{PG}(2m-1, 3) \setminus \mathcal{S}$ with secant intersection numbers $r(P) = 1, 2, 3$, respectively.
\begin{itemize}
\item \textbf{Type 1:} $P \in \mathcal{S}$, which corresponds to the number of such codimension-one subcodes that $N_1 = |\mathcal{S}| = 3^m - 1$.
\item \textbf{Types 2, 3, 4:} $P \notin \mathcal{S}$, which are classified according to their secant intersection numbers $r(P) \in \{1, 2, 3\}$.
\end{itemize}

Denote $M_r = \#\{P \in \mathrm{PG}(2m-1, 3) \setminus \mathcal{S} \mid r(P) = r\}$ for $r \in \{1, 2, 3\}$.
In order to  determine the frequency $M_r$, we establish a system of three equations as follows, i.e., (\ref{eq:ext_total}), (\ref{eq:secant_sum1}), and (\ref{eq:secant_sum2}).
\begin{enumerate}
\item
%\textbf{Total number of external points}
Note that the total number of projective points in $\mathrm{PG}(2m-1, 3)$ is
$$\frac{3^{2m}-1}{2} = \frac{q^2-1}{2}.$$
By definitions, we get
\begin{equation}\label{eq:ext_total}
M_1 + M_2 + M_3 = \frac{q^2 - 1}{2} -N_1 =\frac{q^2 - 1}{2} -(3^m-1) = \frac{(3^m-1)^2}{2}= \frac{n^2}{2} .
\end{equation}

\item
%\textbf{First Moment of Secant Incidences:}
Note that the cap $\mathcal{S}$ determines $\binom{n}{2}$ secant lines. Since each projective line over $\mathbb{F}_3$ contains $3+1=4$ points, and $\mathcal{S}$ is a cap (no three points are collinear), every secant line contains exactly $2$ internal points in $\mathcal{S}$ and $4-2=2$ external points in $\mathrm{PG}(2m-1, 3) \setminus \mathcal{S}$. Counting the incident pairs $(P, \ell)$ yields
\begin{equation}\label{eq:secant_sum1}
M_1 + 2M_2 + 3M_3 = 2 \binom{n}{2} = n(n-1) = (3^m - 1)(3^m - 2),
\end{equation}
where $P \in \ell \setminus \mathcal{S}$ and $\ell$ is a secant line.
\item
%\textbf{Second Moment of Secant Incidences via Dual Weight 4:}
Note that $d(\mathcal{C}^\perp)=4$. Applying the fifth identity of (\ref{eq:PPM}) to the known weight distribution of $\mathcal{C}$, the number of codewords of weight $4$ in the dual code $\mathcal{C}^\perp$ is given by
\begin{equation}\label{eq:A4_dual}
A_4(\mathcal{C}^\perp) = \frac{5n(n-2)}{12}.
\end{equation}

Since $\mathbb{F}_3^* = \{\pm 1\}$, the codewords of weight $4$ in $\mathcal{C}^\perp$ appear in pairs $\{\mathbf{c}, -\mathbf{c}\}$ sharing the same support, which implies that there are $A_4(\mathcal{C}^\perp)/2$ minimally linearly dependent $4$-subsets of $\mathcal{S}$. In $\mathrm{PG}(2m-1, 3)$, every such $4$-subset constitutes a complete quadrangle and thus gives rise to $3$ diagonal points, each defined by an intersection of two secant lines. Conversely, any pair of distinct secants meeting at an external point $P$ determines such a dependent $4$-subset. Double counting the pairs of intersecting secants at external points gives

%Since each projective weight-4 codeword in $\mathcal{C}^\perp$ corresponds to a minimal linear dependence among 4 points of $\mathcal{S}$ (yielding 3 pairs of intersecting secant lines), and conversely, any pair of distinct secants meeting at an external point $P$ determines such a weight-4 configuration, double counting the pairs of intersecting secants at external points gives:
\begin{equation}\label{eq:secant_sum2}
\sum_{P \notin \mathcal{S}} \binom{r(P)}{2} = M_2 + 3M_3 = 3\cdot  \frac{A_4(\mathcal{C}^\perp)}{2} = \frac{5n(n-2)}{8} = \frac{5(3^m-1)(3^m-3)}{8}.
\end{equation}
\end{enumerate}

Solving the linear system of equations (\ref{eq:ext_total}), (\ref{eq:secant_sum1}) and (\ref{eq:secant_sum2}) uniquely gives
\begin{align*}
M_1 &= \frac{n(n+6)}{8} = \frac{(3^m-1)(3^m+5)}{8}, \\
M_2 &= \frac{n(n-2)}{4} = \frac{3(3^m-1)(3^{m-1}-1)}{4}, \\
M_3 &= \frac{n(n-2)}{8} = \frac{(3^m-1)(3^m-3)}{8}.
\end{align*}
The distributions of $A_{w_i}(\mathcal{D})$ in Table~\ref{tab:subcode-freq-all} follow from Theorem~\ref{thm:sub-distribution} and the values of these frequencies $M_r$ and $N_1$.
\end{proof}

\begin{example}
Some examples on the frequency numbers obtained by the proof of Theorem \ref{thm:subcode-distribution-corrected} are given in Table \ref{tab:subcode_examples_mk}, which is confirmed by Magma programs.
\begin{table}[H]
 \centering \caption{Some examples on the frequency numbers for given $(m, k')$} \label{tab:subcode_examples_mk}
  \renewcommand{\arraystretch}{1.3}
  \begin{tabular}{cccccc}
  \hline $m$ & $k'$ & $n$ & $M_1$ & $M_2$ & $M_3$ \\
  \hline 3 & 2 & 26 & 104 & 156 & 78 \\
  5 & 2 & 242 & 7502 & 14520 & 7260 \\
  5 & 4 & 242 & 7502 & 14520 & 7260 \\
  7 & 2 & 2186 & 598964 & 1193556 & 596778 \\
   \hline \end{tabular}
   \end{table}
\end{example}

\begin{example}
Let $(m,k')=(3,2)$. The frequency distribution of $A_{w_i}(\mathcal{D})$ over all codimension-one subcodes of $\mathcal{C}$ in Theorem \ref{thm:subcode-distribution-corrected}
is given in Table \ref{mainre}, which is confirmed by Magma programs.
\begin{table}[H]
\centering \caption{The frequency distribution of $A_{w_i}(\mathcal{D})$ for $(m, k') = (3, 2)$}\label{mainre}
\renewcommand{\arraystretch}{1.4}
\begin{tabular}{ccccc} \hline
Type & $A_{w_1}(\mathcal{D})$ & $A_{w_2}(\mathcal{D})$ & $A_{w_3}(\mathcal{D})$ & Number of $\mathcal{D}$ \\
\hline 1 & 132 & 80 & 30 & 26 \\
2 & 96 & 98 & 48 & 104 \\
 3 & 102 & 86 & 54 & 156 \\
 4 & 108 & 74 & 60 & 78 \\ \hline
\end{tabular}
\end{table}
\end{example}

\subsection{Three classes of linear codes with four non-zero weights}

In this subsection, we obtain three classes of linear codes \(\mathcal C^*(w_i)\) with four non-zero weights and determine the parameters of these codes.

%Specifically, for each such $\mathcal B$, the corresponding codeword in $\mathcal C^*(w_i)$ has weight
%\[\wt(\mathcal D_{\mathcal B}) = A_{w_i} - A_{w_i}(\mathcal B).\]

Before giving and proving the main results of this subsection, we firstly give the auxiliary results in Lemmas \ref{lem:C*-dist-r1} and \ref{lem:dim-cyclic-proof}, which will be needed in proving the main results of Theorem \ref{thm:main}.

For a codimension-one subcode $\mathcal D$ of $\mathcal C$, let $\mathbf{c}_{\mathcal D}$ denote the codeword in $\mathcal C^*(w)$ corresponding to any linear functional whose kernel is $\mathcal D$. From Corollary $7$ of \cite{Huang2026} with $r=1$, the following results in Lemma \ref{lem:C*-dist-r1} is clear and we omit its detailed proof.

%will play an important role for determining  the parameters of \(\mathcal C^*(w_i)\).

\begin{lemma}\label{lem:C*-dist-r1}
Let $\mathcal C$ be an $[n,k]$ linear code over $\F_q$, and assume that
$\dim(\mathcal C^*(w)) = \dim(\mathcal C)$.
Then the weight distribution of $\mathcal C^*(w)$ with $A_w \ne 0$ is determined by the distribution of $A_w(\mathcal D)$ over all codimension-one subcodes $\mathcal D$ of $\mathcal C$. Specifically:
\begin{enumerate}
    \item The minimum distance of $\mathcal C^*(w)$ is
    \[
    d(\mathcal C^*(w)) = A_w(\mathcal C) - \max_{\mathcal D} A_w(\mathcal D),
    \]
    where $\mathcal D$ ranges over all codimension-one subcodes of $\mathcal C$.

    \item Each codimension-one subcode $\mathcal{D}$ of $\mathcal{C}$ corresponds to $q-1$ non-zero codewords in $\mathcal{C}^*(w)$, each codeword $\mathbf{c}_{\mathcal D}$ in $\mathcal C^*(w)$ has Hamming weight
    \[
    wt(\mathbf{c}_{\mathcal D})= A_w(\mathcal C) - A_w(\mathcal D).
    \]
    Then the weight distribution of $\mathcal C^*(w)$ is given by $ A_0(\mathcal C^*(w))=1$  and
    \[
    A_i(\mathcal C^*(w)) = (q-1)\#\{\mathcal D : A_w(\mathcal D) = A_w(\mathcal C) - i\},
    \]
    for $i\geq1$, where $\mathcal D$ ranges over all codimension-one subcodes of $\mathcal C$.
\end{enumerate}
\end{lemma}

\begin{lemma}\label{lem:dim-cyclic-proof}

Let $m \geq 3$ be an odd integer, $q=3^m$, $k' \geq 2$ be an even integer with $\gcd(m,k')=1$. Let $\mathcal{C}$ be the ternary $[q-1, 2m]$ cyclic code defined by (\ref{cyclic}).
For each non-zero weight $w \in \{w_1, w_2, w_3\}$ with $A_{w} \ne 0$,  we have $\dim(\mathcal C^*(w)) = \dim(\mathcal C)=2m$.
\end{lemma}

\begin{proof}
Let $L(\mathcal{C})$ denote the dual space of all linear functionals on $\mathcal{C}$, which satisfies $\dim(L(\mathcal{C})) = \dim(\mathcal{C}) = 2m$. For each non-zero weight $w \in \{w_1, w_2, w_3\}$, we define the map
\[
\phi \colon L(\mathcal{C}) \to \mathbb{F}_3^{A_w}, \quad \ell \mapsto \big(\ell(\mathbf{c})\big)_{\mathbf{c} \in \wt^{-1}(w)}.
\]
By definition, the  code $\mathcal{C}^*(w)$ is the image of $\phi$, that is,
\[
\mathcal{C}^*(w) = \im\phi \cong L(\mathcal{C})/\ker\phi.
\]

A linear functional $\ell \in L(\mathcal{C})$ lies in $\ker\phi$ if and only if $\ell(\mathbf{c}) = 0$ for all $\mathbf{c} \in \wt^{-1}(w)$. This is equivalent to saying that $\wt^{-1}(w) \subseteq \ker\ell$.

Suppose that there exists a non-zero functional $\ell \in \ker\phi$. Then $\mathcal{D} = \ker\ell$ is a codimension-one subcode of $\mathcal{C}$. The containment $\wt^{-1}(w) \subseteq \mathcal{D}$, together with the fact that $\mathcal{D} \subseteq \mathcal{C}$, implies that $\mathcal{D}$ contains all codewords of weight $w$ in $\mathcal{C}$, which means
\[
A_w(\mathcal{D}) = A_w(\mathcal{C}).
\]
However, this contradicts Theorem~\ref{thm:subcode-distribution-corrected}, which shows that $A_w(\mathcal{D}) < A_w(\mathcal{C})$ for every codimension-one subcode $\mathcal{D}$ of $\mathcal{C}$.
Therefore, no such non-zero functional exists, which implies $\ker\phi = \{\mathbf{0}\}$. Hence, the map $\phi$ is injective, and we conclude that
\[
\dim(\mathcal{C}^*(w)) = \dim(\im\phi) = \dim(L(\mathcal{C})) = 2m.
\]
This completes the proof.
\end{proof}

\begin{theorem}\label{thm:main}
Let the symbols and notations be as above. Let $\mathcal{C}^*(w_i)$ be the code obtained by applying the weight-function construction in (\ref{eq:C*}) to the cyclic code $\mathcal{C}$ at weight $w_i$ for $i \in \{1, 2, 3\}$.
Then $\mathcal{C}^*(w_i)$ is an $[A_{w_i}, 2m, d_{w_i}]$ ternary linear code with weight distributions given in Tables~\ref{tab:new1}, \ref{tab:new2}, and \ref{tab:new3}, where $A_{w_i} = A_{w_i}(\mathcal{C})$ is given in (\ref{eq3}) and

\begin{eqnarray*}
d_{w_i} =
\left\{ \begin{array}{ll}
     3^{m-1}(2\cdot 3^{m-1} +3^{(m-1)/2} -1) ;                                        & \mbox{ if $i=1$,} \\
       2\cdot 3^{2m-2}             ;                                                  & \mbox{ if $i=2$,} \\
          2\cdot 3^{2m-2} - 2\cdot 3^{\frac{3m-3}{2}} -4\cdot 3^{m-2}       ;          & \mbox{ if $i=3$.}
\end{array}
\right.
\end{eqnarray*}

\begin{table}[H]
\centering \caption{The weight distribution of $\mathcal{C}^*(w_1)$}\label{tab:new1}
\renewcommand{\arraystretch}{1.5}
\begin{tabular}{cc}
\hline
Weight  &  Multiplicity \\ \hline
$0$ & $1$ \\
$3^{m-1}(2\cdot 3^{m-1} +3^{(m-1)/2} -1          )$ & $2(3^m-1)$ \\
$   2\cdot 3^{m-1} ( 3^{m-1} +3^{(m-1)/2}              )         $ & $\frac{(3^m-1)(3^m+5)}{4}$ \\
$2\cdot (3^{2m-2} + 3^{(3m-3)/2} - 3^{m-2}     )$ & $\frac{3(3^m-1)(3^{m-1}-1)}{2}$ \\
$2\cdot (3^{2m-2} + 3^{(3m-3)/2} - 2\cdot 3^{m-2}   )$ & $\frac{(3^m-1)(3^m-3)}{4}$ \\ \hline
\end{tabular}
\end{table}

\begin{table}[H]
\centering \caption{The weight distribution of $\mathcal{C}^*(w_2)$}\label{tab:new2}
\renewcommand{\arraystretch}{1.5}
\begin{tabular}{cc} \hline
Weight  &  Multiplicity \\ \hline
$0$ & $1$ \\
$2\cdot 3^{2m-2}+ 2\cdot 3^{m-1}$                  & $2(3^m-1)$ \\
$2\cdot 3^{2m-2}$                                  & $\frac{(3^m-1)(3^m+5)}{4}$ \\
$2\cdot 3^{2m-2} + 4\cdot 3^{m-2}$                  & $\frac{3(3^m-1)(3^{m-1}-1)}{2}$ \\
$2\cdot 3^{2m-2} + 8\cdot 3^{m-2}$                  & $\frac{(3^m-1)(3^m-3)}{4}$ \\ \hline
\end{tabular}
\end{table}
\begin{table}[H]
\centering \caption{The weight distribution of $\mathcal{C}^*(w_3)$}\label{tab:new3}
\renewcommand{\arraystretch}{1.5}
\begin{tabular}{cc} \hline
Weight  &  Multiplicity \\
\hline $0$ & $1$ \\
$2\cdot 3^{2m-2}- 3^{(3m-3)/2}- 3^{m-1}     $                  & $2(3^m-1)$ \\
$2\cdot 3^{2m-2} - 2\cdot 3^{\frac{3m-3}{2}}$                  & $\frac{(3^m-1)(3^m+5)}{4}$ \\
$2\cdot 3^{2m-2} - 2\cdot 3^{\frac{3m-3}{2}} -2\cdot 3^{m-2}$                 & $\frac{3(3^m-1)(3^{m-1}-1)}{2}$ \\
$2\cdot 3^{2m-2} - 2\cdot 3^{\frac{3m-3}{2}} -4\cdot 3^{m-2}$     & $\frac{(3^m-1)(3^m-3)}{4}$ \\
\hline
\end{tabular}
\end{table}

\end{theorem}

\begin{proof}
By definitions, the code length of $\mathcal{C}^*(w_i)$ is $A_{w_i}(\mathcal{C})$.
%From Lemma~\ref{lem:dim-cyclic-proof}, we have $\dim(\mathcal{C}^*(w_i)) = \dim(\mathcal{C}) = 2m$.
By Lemma \ref{lem:C*-dist-r1}, the conclusions then follow from Equations (\ref{eq2}) and (\ref{eq3}), Lemma~\ref{lem:dim-cyclic-proof} and Theorem \ref{thm:subcode-distribution-corrected}. This completes the proof.
%Next we prove the conclusions on the weight distribution of $\mathcal{C}^*(w_i)$. We only give the proof for $w_1$. The proofs of the other conclusions for $w_2$ and $ w_3$ are similar to that of $w_1$ and thus omitted.
\end{proof}

\begin{remark}
In Theorem~\ref{thm:main}, every nonzero weight in $\mathcal{C}^*(w_i)$ ($i=1,2,3$) is a multiple of $3$. Thus, from Lemma \ref{lem-judge} we have that these three classes of linear codes $\mathcal{C}^*(w_i)$ are self-orthogonal. Furthermore, by the Ashikhmin--Barg criterion, these codes are minimal for any odd integer $m \ge 3$ since $w_{\min}/w_{\max} > 2/3$, where $w_{\min}$ and $w_{\max} $ are the minimal nonzero weight and the maximal nonzero weight of $\mathcal{C}^*(w_i)$ in Theorem \ref{thm:main}, respectively.
\end{remark}

%\begin{proof} By definition, the code length of $\mathcal{C}^*(w_i)$ is equal to $A_{w_i}(\mathcal{C})$. According to Lemma~\ref{lem:dim-cyclic-proof}, we have $\dim(\mathcal{C}^*(w_i)) = \dim(\mathcal{C}) = 2m$. By Lemma~\ref{lem:C*-dist-r1}, each codimension-one subcode $\mathcal{D}$ of $\mathcal{C}$ corresponds to $3 - 1 = 2$ codewords in $\mathcal{C}^*(w_i)$ with Hamming weight $A_{w_i}(\mathcal{C}) - A_{w_i}(\mathcal{D})$. We illustrate the computation for $\mathcal{C}^*(w_1)$. Combining the value of $A_{w_1}(\mathcal{C})$ from (\ref{eq3}) and the values of $A_{w_1}(\mathcal{D})$ from Theorem~\ref{thm:subcode-distribution-corrected} (Table~\ref{tab:subcode-freq-all}), we have: \begin{itemize} \item For Type~1 subcodes $\mathcal{D}_1$, $\mathrm{wt}(c) = A_{w_1}(\mathcal{C}) - A_{w_1}(\mathcal{D}_1) = 2 \cdot 3^{2m-2} + 3^{\frac{m-1}{2}} - 3^{m-2}$, yielding $2(3^m - 1)$ codewords; \item For Type~$j$ subcodes $\mathcal{D}_j$ ($j = 2, 3, 4$), the corresponding weights and numbers of codewords are similarly obtained by $A_{w_1}(\mathcal{C}) - A_{w_1}(\mathcal{D}_j)$ and $2 \times (\text{frequency of Type } j)$. \end{itemize} This establishes the weight distribution of $\mathcal{C}^*(w_1)$ in Table~\ref{tab:new1}. The weight distributions of $\mathcal{C}^*(w_2)$ and $\mathcal{C}^*(w_3)$ can be derived completely analogously. Note that for $\mathcal{C}^*(w_2)$, the subcodes of Type~1 and Type~3 yield the identical weight $2 \cdot 3^{2m-2} + 2 \cdot 3^{m-2}$, which combines into a single frequency of $\frac{(3^m-1)(3^m+1)}{2}$. The proof is thus complete. \end{proof}

\begin{example}
Let $m=3$ and $k'=2$. Then the cyclic code $\mathcal{C}$ defined by (\ref{cyclic}) is a $[26,6,15]$ ternary code with weight enumerator
$$
1+312 x^{15}+260 x^{18}+ 156 x^{21}.
$$
Theorem \ref{thm:main} gives the following three codes:
\begin{itemize}
    \item $\mathcal{C}^*(15)$ is a $[312, 6, 180]$ ternary linear code with weight enumerator
    $$
    1 + 52x^{180} + 156x^{204} + 312x^{210} + 208 x^{216}  .
    $$

    \item $\mathcal{C}^*(18)$ is a $[260, 6, 162]$ ternary linear code with weight enumerator
    $$
    1 + 208x^{162} + 312 x^{174}+ 52x^{180} + 156x^{186}.
    $$

    \item $\mathcal{C}^*(21)$ is a $[156, 6, 96]$ ternary linear code with weight enumerator
   $$
    1 + 156x^{96} + 312x^{102} + 208 x^{108} + 52 x^{126}.
   $$
   %This code is optimal.

\end{itemize}
\end{example}

\section{Conclusion}\label{sec-summary}

In this paper, we have investigated the construction of linear codes via weight functions proposed by Huang et al. \cite{Huang2026} and extended this method beyond the two-weight setting. Based on a well-known family of ternary three-weight cyclic codes, we analyzed the relationship between the fixed-weight codeword sets and their codimension-one subcodes, which led to three new infinite families of ternary linear codes with few nonzero weights. Furthermore, the parameters of these codes are explicitly determined and these codes are shown to be self-orthogonal and minimal.

It is worth pointing out that although we focused on the known cyclic code associated with a specific monomial planar function to demonstrate the explicit parameters in this paper, the theoretical framework and geometric analysis employed here are potentially applicable to other settings. Specifically, for any cyclic code $\mathcal{C}$ whose associated projective point set takes the form of \eqref{SS} with $f(x)$ being any monomial planar function satisfying Lemma~\ref{lem:planarity}, if $\mathcal{C}$ has three nonzero weights and its dual code has  minimum distance $d(\C^\perp) \ge 4$, we conjecture that similar infinite families of linear codes $\C^*(w)$ can be derived and their parameters can be determined via an analogous approach.
An interesting direction for future research would be to explore whether this subcode-analysis approach and the weight-function framework can be generalized to linear codes derived from other functions or other families of linear codes with few weights over general finite fields $\mathbb{F}_q$.

%\section*{Acknowledgements}This paper was supported by the Basic and Applied Basic Research Foundation of Guangdong Province of China under Grant number 2026A1515011223, the National Natural Science Foundation of China under grant numbers 12671651 and 12231015, and  the Science and Technology Projects of Xizang Autonomous Region of China under grant number XZ202502JD0036.

\end{document}